\documentclass[12pt]{article}
\pdfoutput=1
\usepackage{amsmath,amssymb}
\usepackage{graphicx}
\usepackage{cite,fancyhdr}

\newcommand{\Slash}[1]{{\ooalign{\hfil#1\hfil\crcr\raise.167ex\hbox{/}}}}

\newcommand{\beq}{\begin{equation}}  \newcommand{\eeq}{\end{equation}}
\newcommand{\bef}{\begin{figure}}  \newcommand{\eef}{\end{figure}}
\newcommand{\bec}{\begin{center}}  \newcommand{\eec}{\end{center}}
\newcommand{\non}{\nonumber}  
\newcommand{\laq}[1]{\label{eq:#1}}

\newcommand{\eq}[1]{(\ref{eq:#1})}

\newcommand{\lac}[1]{\label{chap:#1}}

\def\({\left(}
\def\){\right)}

\def\O{\mathcal{O}}

\newcommand{\AND}{~{\rm and}~}

\newcommand{\GEV}{ {\rm ~GeV} }
\newcommand{\TEV}{ {\rm ~TeV} }

\def\a{\alpha}
\def\b{\beta}

\def\d{\delta}

\def\g{\gamma}

\def\m{\mu}

\def\s{\sigma}

\def\D{\Delta}

\def\tl{\tilde}
\def\*{\dagger}
\usepackage[affil-it]{authblk}
\usepackage[left=2cm,right=2cm,top=2cm,bottom=2cm,a4paper]{geometry}

\begin{document}
\title{\bf \large 
Muon $g-2$ in Higgs-anomaly mediation
}

\author[1,2]{{\normalsize  Tsutomu T. Yanagida}}
\author[3]{{\normalsize Wen Yin}}
\author[4]{{\normalsize Norimi Yokozaki}}

\affil[1]{\small 
T.D.Lee Institute and School of Physics and Astronomy,
Shanghai Jiao Tong University, 

Shanghai 200240, China}
\affil[2]{\small  Kavli IPMU (WPI), UTIAS,
The University of Tokyo,  Kashiwa, Chiba 277-8583, Japan}

\affil[3]{\small 
Department of Physics, KAIST, Daejeon 34141, Korea}

\affil[4]{\small 
Department of Physics, Tohoku University,  

Sendai, Miyagi 980-8578, Japan}

\date{}

\maketitle

\renewcommand{\headrulewidth}{0pt}

\begin{abstract}
\noindent
A simple model for the explanation of the muon anomalous magnetic moment was proposed by the present authors within the context of the minimal supersymmetric standard model~\cite{Yin:2016shg, Yanagida:2016kag}:~{\it Higgs-anomaly mediation}. In the setup, squarks, sleptons, and gauginos are massless at tree-level, but the Higgs doublets get large negative soft supersymmetry (SUSY) breaking masses squared $m_{H_u}^2 \simeq m_{H_d}^2 < 0$ at a certain energy scale, $M_{\rm inp}$. The sfermion masses are radiatively generated by anomaly mediation and Higgs-loop effects, and gaugino masses are solely determined by anomaly mediation. Consequently, the smuons and bino are light enough to explain the muon $g-2$ anomaly while the third generation sfermions are heavy enough to explain the observed Higgs boson mass. The scenario avoids the SUSY flavor problem as well as various cosmological problems, and is consistent with the radiative electroweak symmetry breaking. In this paper, we show that, although the muon $g-2$ explanation in originally proposed Higgs-anomaly mediation with $M_{\rm inp}\sim 10^{16}$\,GeV is slightly disfavored by the latest LHC data, the muon $g-2$ can still be explained at $1\sigma$ level when Higgs mediation becomes important at the intermediate scale, $M_{\rm inp} \sim 10^{12}\GEV$. The scenario predicts light SUSY particles that can be fully covered by the LHC and future collider experiments. We also provide a simple realization of $m_{H_u}^2 \simeq m_{H_d}^2 < 0$ at the intermediate scale.
\end{abstract}

\clearpage
\section{Introduction}

The discrepancy of the muon anomalous magnetic moment ($g-2$) is one of the important indications for the existence of new physics that couples to the standard model (SM). 
The discrepancy is given by
\beq
\laq{g2m}
\D a_\mu = a_\mu^{\rm EXP} - a_\mu^{\rm SM} = (27.4 \pm 7.3) \times 10^{-10},
\eeq
where $a_\m^{\rm SM}$ is the SM prediction of the muon $g-2$~\cite{Keshavarzi:2018mgv}~(see also Refs.\,\cite{Davier:2017zfy,Keshavarzi:2019abf}), and 
$a_\mu^{\rm EXP}$ is the experimental value~\cite{Bennett:2006fi,Roberts:2010cj}.
Importantly, new physics models to explain the muon $g-2$ anomaly should guarantee the suppression of flavor-changing neutral current (FCNC) processes and avoidance of cosmological problems.

Among the new physics models, the minimal supersymmetric (SUSY) extension of the SM (MSSM) is one of the promising candidates since it also provides a solution to the (large) hierarchy problem and unification of the fundamental forces. To avoids the SUSY FCNC problem, masses for squarks and sleptons should not be dominantly generated by gravity mediation, which is believed to induce the FCNC. 
The simplest possibility is to assume that the squarks and slepton are massless at the high energy scale and their masses are generated by flavor-safe mediation mechanisms of SUSY breaking. 
It is known that sequestering between the SUSY breaking sector and visible sector in various contexts~\cite{Inoue:1991rk, 
Randall:1998uk,Nelson:2000sn, Luty:2001jh} or Nambu-Goldstone (NG) hypothesis of the sfermions~\cite{Kugo:1983ai, Yanagida:1985jc} can suppress the dangerous gravity mediated sfermion masses.\footnote{ 
One can also consider other flavor-safe mediation mechanisms e.g. Refs.~\cite{Dine:1993yw, Dine:1994vc, Dine:1995ag, Kaplan:1999ac, Chacko:1999mi}.} 
In addition, to avoid the cosmological disaster, the Polonyi problem, the SUSY breaking field, $Z$, should be charged under some symmetry with a suppressed vacuum expectation value (VEV).\footnote{This kind of cosmological safety with alleviation of the gravitino problem with $m_{3/2}\gtrsim \O(10)\TEV$ can be found in the pure-gravity mediation scenario~\cite{Ibe:2006de,Ibe:2011aa}, minimal-split SUSY~\cite{ArkaniHamed:2012gw} or the split-SUSY~\cite{ArkaniHamed:2004yi,Giudice:2004tc}. }
Then, the gaugino masses can be only generated by anomaly mediation~\cite{Randall:1998uk, Giudice:1998xp} or gauge mediation. In the latter case, it is known that the $g-2$ discrepancy is difficult to be explained in a simple setup (see e.g. Ref.~\cite{Bhattacharyya:2018inr}). 
Therefore, we consider the sequestering scenario with anomaly mediation.\footnote{In the NG hypothesis of the sfermions the muon $g-2$ anomaly can be also explained in a similar manner, but one should take into account the sigma-model anomaly mediation if the NG modes arise from a compact K\"{a}hler manifold~\cite{Yanagida:2019evh}: e.g. the gaugino spectra are different from those predicted in anomaly mediation. If the low energy NG-modes do not induce the sigma-model anomaly mediation in some setup the conclusion is the same as this paper.} 
Note that the tachyonic slepton problem can be solved as we will discuss soon.

One of the simplest sequestering scenarios to explain the muon $g-2$ anomaly is known to be Higgs-anomaly mediation~\cite{Yin:2016shg, Yanagida:2016kag}, where 
quark and lepton multiplets are sequestered from the SUSY breaking sector, 
but Higgs doublets couple to the SUSY breaking field directly. 
Then, at the renormalization group (RG) scale,  $\mu_{\rm RG}=M_{\rm inp}$, the SUSY breaking masses for the Higgs doublets are generated as\footnote{This condition is needed, otherwise the non-vanishing $D$-term leads to the splitting of the light spectrum and cause the tachyonic scalars by one-loop effects. We will provide a simple realization of the condition in Sec.~\ref{Sec:4}.} 
\begin{equation}\non
m_{H_u}^2=m_{H_d}^2=-c_h m_{3/2}^2, 
\end{equation}
where $c_h$ is assumed to be positive. At the tree level, the other soft SUSY breaking parameters are   
\begin{equation}
\laq{gaugino}\non
M_1=M_2=M_3=0,\end{equation}
\begin{equation}
\non
~ {A}_u={A}_d={A}_e=0,~\end{equation}\begin{equation}
\laq{HM}
{m}_{\tl{ Q}}^2={m}_{\tl{ u}}^2={m}_{\tl{ d}}^2={m}_{\tl{ L}}^2={m}_{\tl{ e}}^2=0.
\end{equation}
We notice again that the SUSY breaking field $Z$ is charged under some symmetry and thus one can not have tree-level $A$-terms and gaugino mass terms. Above $M_{\rm inp}$, renormalization group runnings of the gaugino masses, the scalar trilinear couplings and sfermion masses follow the anomaly mediation trajectories. Below $M_{\rm inp}$, sfermion masses also feel Higgs-loop effects. 

In summary, we define Higgs-anomaly mediation with five free parameters: the coupling between $Z$ and the Higgs doublets, the gravitino mass, the ratio of VEVs of the up and down type Higgs doublets, the sign of the Higgsino mass parameter and the scale to define $m_{H_u}^2=m_{H_d}^2=-c_h m_{3/2}^2$. 
\beq
\laq{freepara}
c_h(>0),~ m_{3/2}, ~{\rm sign}{\mu}, ~\tan\b, ~M_{\rm inp}.
\eeq
The parameters in the Higgs sector, $|\mu| \AND B\mu$, are determined by the conditions of the electroweak symmetry breaking (EWSB), once the five parameters are given. Here and hereafter, we assume the CP symmetry is only violated by the Yukawa interactions of SM. Then, $\m$ and $B\m$ are real parameters.

Although the sfermions are massless at the tree-level, radiative corrections lead to broad SUSY mass spectra. 
It has been shown that masses for third generation sfermions are significantly enhanced
due to the large Yukawa coupling of $\mathcal{O}(1)$ from Higgs-loops at the one-loop level, when $m_{H_{u,d}}^2$ are negative and large~\cite{Yamaguchi:2016oqz}. The Higgs-loop effects with negative $m_{H_{u,d}}^2$ is called Higgs mediation.\footnote{See Refs.~\cite{Yin:2016shg,Yanagida:2018eho,Cox:2018vsv, Endo:2019bcj,
Badziak:2019gaf, Yanagida:2019evh} for applications of Higgs mediation.} 
With the heavy stops, the observed Higgs boson mass of 125\,GeV is consistently explained from stop loops~\cite{Okada:1990vk,
Ellis:1990nz,
Haber:1990aw,
Okada:1990gg,
Ellis:1991zd}. 
On the other hand, Higgs loop effects at the one-loop level to the masses for squarks and sleptons of the first two generations are small due to the tiny Yukawa couplings. The masses are essentially determined by anomaly mediation and Higgs mediation at the two-loop level~\cite{Yin:2016shg}. Notably, by taking into account the Higgs mediation effects, the slepton masses are not tachyonic anymore. The gaugino masses are solely determined by anomaly mediation.
It was shown that, in a brane-world scenario, where the compactification scale as well as $M_{\rm inp}$ is taken to be $10^{16}$\,GeV, the muon $g-2$ anomaly and the unification of bottom-tau or top-bottom-tau Yukawa couplings can be simultaneously explained~\cite{Yin:2016shg,Yanagida:2018eho}. The light squarks can be checked at the LHC when the wino dark matter is compatible with thermal leptogenesis~\cite{Yanagida:2016kag}. Since the CKM matrix is an only source for flavor violations, the dangerous FCNC processes are suppressed~\cite{Yanagida:2018eho}.

In this paper, we study the $1\s$ explanation of the muon $g-2$ anomaly based on Higgs-anomaly mediation in detail. 
We first review the scenario based on the brane-world where the compactification scale, as well as $M_{\rm inp}$, is taken to be $10^{16}\GEV$. We show that, in this case, the muon $g-2$ discrepancy \eq{g2m} can not be explained at the $1\,\s$ level by taking into account the latest LHC data.
In Sec.\ref{sec:3}, we point out that the muon $g-2$ can be explained at $1\sigma$ level for $M_{\rm inp} \sim 10^{12}\GEV$. 
In Sec.~\ref{Sec:4}, we provide a simple example model to explain $m_{H_u}^2\simeq m_{H_d}^2 <0$ and $10^{12}\GEV \lesssim M_{\rm inp} \lesssim 10^{16}\GEV$. The last section is devoted to discussions and conclusions.

\section{Higgs-anomaly mediation}
\lac{2}

\subsection{Review on a simple explanation of the muon $g-2$}
Let us review the setup of Higgs-anomaly mediation in more detail. 
One of the realizations of Higgs-anomaly mediation is a brane-world scenario, where quarks and leptons live in one brane (matter brane). The matter brane is geometrically separated from the other brane (SUSY-breaking brane), where the SUSY breaking field $Z$ lives. The important thing is that the Higgs multiplets live in the bulk. 
The K\"{a}hler potential then takes~\cite{Inoue:1991rk, 
Randall:1998uk}
\begin{eqnarray}
K = - 3 M_{P}^2 \ln \left[1 - \frac{f(Z, Z^\dag) + \phi_i^\dag \phi_i + \Delta K}{3 M_{P}^2} \right], 
\end{eqnarray}
where $M_P\simeq 2.4\times 10^{18}\GEV$ is the reduced Planck mass, and $\phi_i$ is a MSSM chiral superfield. 
The $F$-term of $Z$, $F_Z$, breaks SUSY spontaneously with $|F_Z| \simeq\sqrt{3}\, m_{3/2} M_P.$
Here, $\Delta K$ contains direct couplings of the Higgs multiplets to the SUSY breaking field $Z$:
\begin{eqnarray}
\Delta K = c_h \frac{|Z|^2}{M_P^2} (|H_u|^2 + |H_d|^2)+(c_\mu+c_b \frac{|Z|^2}{M_P^2}) H_u H_d+h.c .
\end{eqnarray}
The resulting mass spectra at the tree-level is \eq{HM} at the compactification scale $\sim M_{\rm inp},$ where $\mu= c_\mu m_{3/2} \AND B \mu= -c_b |F_Z|^2/M_P^2+c_\mu m_{3/2}^2$ are derived. Here, the universal coupling $c_h$ is assumed for simplicity. A possible model to explain the universal coupling is given in Sec.~\ref{Sec:4}.
Since the Higgs doublets directly couple to the matter multiplets, the Yukawa interactions can be given in the usual way as
\beq
W\ni -y_u H_u Q u-y_d H_d Q d-y_e H_d L e.
\eeq
This is the ultra-violet (UV) realization from the brane-world scenario. 
However, the UV model should not be called a ``UV completion" because the higher dimensional field theory is quite non-trivial to have a UV fixed point. 
Nonetheless, it may be derived from string-theory or M-theory. 
Alternatively, the conformal sequestering should also work~\cite{Chacko:2001jt}. 
If we take the brane-world scenario with compactification scale $1/L= 10^{16}\GEV$ [around the grand unified theory (GUT) scale], the input scale should be 
\beq
M_{\rm inp}\simeq \frac{1}{L}=10^{16}\GEV,
\eeq
below which the four dimensional description starts and SUSY breaking mediation from the Higgs doublets to the sfermion masses becomes important. 
The compactification scale of around $10^{16}$\,GeV is chosen from the following considerations: a) $M_{\rm inp}$ may not be too far away from $M_{P}$ from the view-point of the UV theory, b) the gauge coupling unification and proton stability  the scale is above $10^{16}\GEV$,
c) the scale should not be too high otherwise the breaking of the sequestering of order $(16\pi^2  M_{P}^2 L^2)^{-1} \times |Z|^2\phi^\*_i\phi_j/M_P^2$ would induce sizable FCNC processes~(see Ref.~\cite{Yanagida:2018eho} and the references therein). 
This setup leads to \eq{HM} at $M_{\rm inp}=10^{16}\GEV$.

The sfermion and gaugino masses are generated by the aforementioned quantum corrections. 
The mass of  sfermion, $X$, is composed of two sources of radiative corrections: anomaly mediation, \beq\laq{AMmat} \d^{\rm AM}{m}_{{X}}^2=- \frac{m^2_{3/2}}{2} \times \frac{d\g_X}{d\log{\m_{\rm RG}}},\eeq and Higgs loop effects, $\d^{\rm HM}{m}_{{X}}^2$:
\beq m_{X}^2=\d^{\rm AM}{ m}_{{X}}^2+\d^{\rm HM}{m}_{{X}}^2,\eeq
where $\g_X$ is the anomalous dimension of sfermion $X$. 
The gaugino mass, on the other hand, is purely given by the anomaly mediation 
\beq
\laq{gaugino}
M_i= m_{3/2}\frac{\beta_i}{g_i},\eeq
where $\beta_i$ is the beta function of the gauge coupling $g_i.$ ($g_1$, $g_2$ and $g_3$ are the gauge couplings of $U(1)_Y$, $SU(2)_L$ and $SU(3)_C$, respectively.)

Anomaly mediation has a notorious tachyonic slepton problem. For instance, 
 a left-handed selectron/smuon acquires a negative mass squared at two-loop level 
\begin{equation}
\laq{Anmd}
 \d^{\rm AM} { m}_{\tl{e}_L, \tl{\m}_L}^2 \sim -\frac{3}{2}\(\frac{g_2^2}{16\pi^2}\)^2  m_{3/2}^2.
\end{equation}
This problem will be solved by taking account of the Higgs loop effects, which is same order but positive~\cite{Yin:2016shg}. 
The negative contributions decrease the smuon masses, enhancing the muon $g-2$.
An important property is that anomaly mediation formulae \eq{AMmat} and \eq{gaugino} do not change with different renormalization scale, i.e. the anomaly mediation is UV insensitive.
Thus one can estimate the contribution at e.g. $\m_{\rm RG}=m_{3/2}$ scale by using the formula with the couplings also at this scale.

\paragraph{Higgs mediation}

The contribution from the RG running via Higgs loops is dubbed as Higgs mediation.
For instance, the contribution to a left-handed stop mass is
\begin{align}
\laq{mQ}
\d^{\rm HM}{m}_{{\tl{t}_L}}^2 &\sim \frac{2}{ 16 \pi^2}  \(y_t^2  +y_b^2\) c_h m_{3/2}^2\log{\(\frac{M_{\rm inp}}{m_{3/2}}\)} ,
\end{align}
where $y_t \AND y_b$ are the top and bottom Yukawa couplings, respectively, and we have taken the 
leading logarithmic approximation.  
On the other hand, the one-loop contributions to the first two generation squarks, are proportional to the tiny Yukawa coupling squares, and thus, are highly suppressed. 
For those first two generation squarks the masses are dominantly generated due to the anomaly mediation, 
 \beq \d^{\rm AM} { m}_{\tl{q}}^2 \sim \(\frac{g_3^2 }{ 16\pi^2}\)^2 8 m_{3/2}^2,\eeq
 unless $c_h>\O(0.1)$ which is disfavored in the muon $g-2$ explanation. In summary, we get the mass hierarchy between the first two generation and third generation squarks. 

A similar mass hierarchy holds in the slepton sector: the staus are much heavier than the smuons and 
selectrons. To discuss the slepton masses for the first two generations we need to consider
the Higgs mediation at the two-loop level. For instance, the left-handed smuon/selectron acquires 
\beq
\laq{HM2}
\d^{\rm HM} { m}_{{{\tl{e}_L,\tl{\m}_L}}}^2 \sim \frac{6g_2^4}{( 16 \pi^2)^2}  c_h m_{3/2}^2\log{\(\frac{M_{\rm inp}}{ m_{3/2} }\)} .\\
\eeq
Here, we use again the leading logarithmic approximation. This positive contribution successfully solves the tachyonic slepton problem. 
Another implication is that the smuon mass squares are of the same order of the gaugino mass and are loop suppressed compared with the gravitino mass if $c_h\lesssim\O(1).$

To sum up, although we have the universal sfermion mass conditions at $\m_{\rm RG}=M_{\rm inp}$, the SUSY mass spectrum is splitting at $\m_{\rm RG}\sim m_{3/2}$. This is contrary to the ordinary flavor-safe mediation scenarios, where the low-energy spectra are also almost universal. 
The splitting spectrum is quite natural in some sense because the fermion mass spectrum in SM is known to be splitting. 

\paragraph{Radiative EWSB and muon $g-2$}
This scenario is favored for the explanation of the muon $g-2$ anomaly due to not only the small smuon and gaugino masses but also the large $|\mu|\tan\b.$
Since $m_{H_u}^2$ and $m_{H_d}^2$ are negative and large, the EWSB requires a large $\mu$-term to have correct electroweak scale:
$|m_{H_u}^2+\mu^2| \ll |m^2_{H_u}|$ which says at a rough estimation, 
\beq 
\laq{mu}
|\mu| \sim |c_h| m_{3/2}. \eeq
This condition can be derived for $\tan\b\gtrsim 1$, and $|m_{ H_u}^2|\gg 100\GEV.$

\begin{figure}[!t]
\begin{center}  
   \includegraphics[width=84mm]{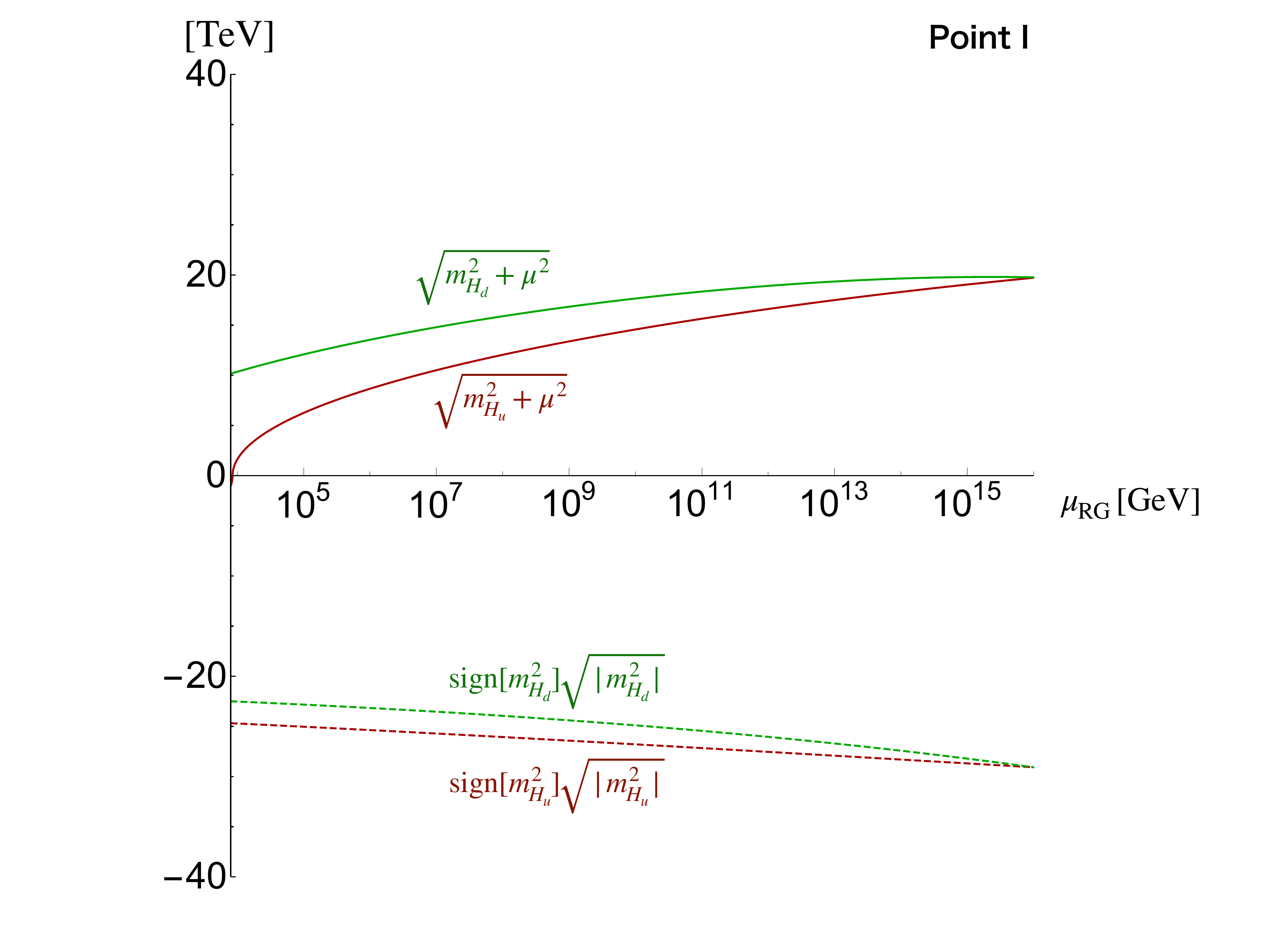}
   \includegraphics[width=84mm]{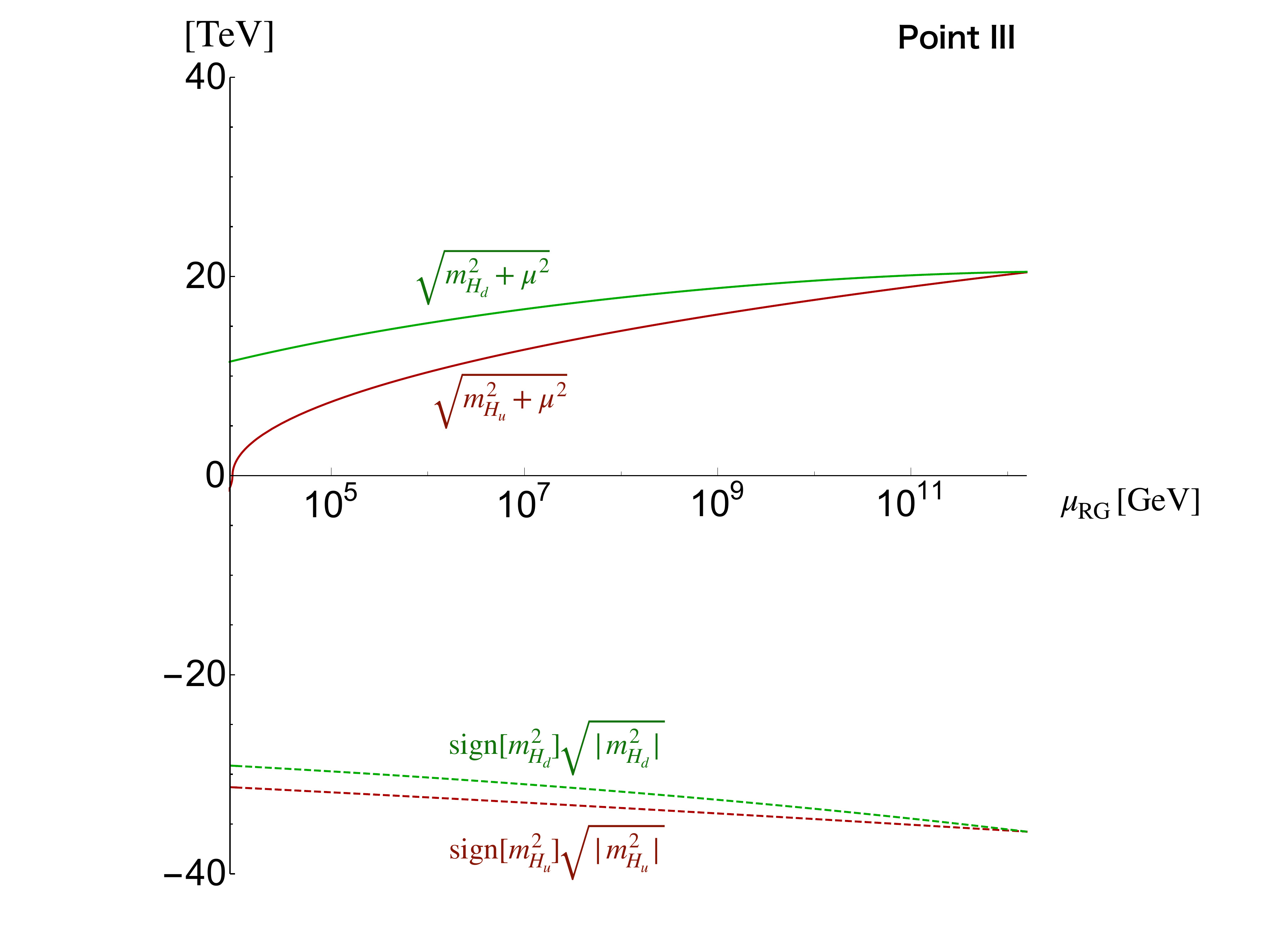}
\end{center}
\caption{The radiative EWSB of Point {\bf I} [left panel] and Point {\bf III} [right panel]. The red and green solid (dashed) lines represent $\sqrt{|m_{H_u}^2+\mu^2|}$ and $\sqrt{|m_{H_d}^2+\mu^2|}$ ($\sqrt{|m_{H_u}^2|}$ and $\sqrt{|m_{H_d}^2|}$), respectively, by varying the renormalization scale.  The sign of the $y$-axis is the sign of the argument. }
\label{fig:EWSB}
\end{figure}

The successful EWSB requires $\tan\b=\O(10)$. This is because a non-tachyonic CP-odd Higgs boson implies
\beq
\laq{mA}
m_A^2\simeq m_{H_d}^2-m_{H_u}^2> 0.
\eeq
Since this condition should be satisfied at $\mu_{\rm RG}\sim m_{3/2}$, the radiative corrections to the SUSY breaking masses for the Higgs doublets are needed to be considered. The radiative corrections are estimated by using the RG equations:
\beq
\frac{d}{d\log\mu}m_{H_u}^2\simeq \frac{1}{16\pi^2} 6y_t^2 m_{H_u}^2,~~ \frac{d}{d\log\mu}m_{H_d}^2\simeq \frac{1}{16\pi^2} (6y_b^2+y_\tau^2) m_{H_d}^2.
\eeq
Given $m^2_{ H_{u}}<0 \AND m^2_{ H_{d}}<0$, one finds that the corrections lead to \eq{mA} only when $6y_b^2+y_\tau^2 \gtrsim 6y_t^2.$ This means $\tan\b=\O(10).$ Note that the large $|\m|\AND \tan\b$ also play an important role in Yukawa coupling unifications~\cite{Yanagida:2019evh,
Yanagida:2018eho,
Yin:2016shg}.
In Fig.\,\ref{fig:EWSB}, the successful radiative EWSB for Points {\bf I} and {\bf III} in Table\,\ref{tab:1} is demonstrated. The detail of analysis as well as data points will be discussed soon. 
In the figure, the RG runnings of the $m_{H_u}^2 +\mu^2$ and $m_{H_d}^2 +\mu^2$, which are the diagonal components of the Higgs mass matrix, are represented. We find that $m_{H_u}^2+\mu^2$ is driven to be negative at around $10\TEV,$ while both of $m_{H_u}^2 +\mu^2$ (Red solid line) and $m_{H_d}^2 +\mu^2$ (green solid line) are positive at higher energy scales. This clearly shows that the EWSB is broken radiatively with a small enough $B\mu$ term. Also, there is no deeper vacuum in the flat direction of the Higgs field $|H_u|=|H_d|$ with vanishing VEVs of the sfermions. For comparison the RG runnings of $m_{H_u}^2$ and $m_{H_d}^2$ are also shown in red and green dashed lines, respectively.

The light smuons and bino with large $|\mu \tan \b|$ enhance the bino-smuon loop contributions to the muon $g-2$ given as~\cite{Lopez:1993vi, Chattopadhyay:1995ae, Moroi:1995yh, Cho:2011rk}
\begin{eqnarray}
(a_\mu)_{\rm SUSY} \simeq \left(
\frac{1 - \delta_{\rm QED}}{1 + \Delta_\mu }
\right) 
\frac{3}{5}\frac{g_1^2}{ 8\pi^2}{ \frac{m_\mu^2  \mu \tan\beta } { M_1^3}}
\,f\left( 
\frac{m_{\tilde{\mu}_L}^2}{M_1^2},
\frac{m_{\tilde{\mu}_R}^2}{M_1^2}
\right) , \label{eq:gm2}
\end{eqnarray}
where $m_{\mu}$ is the muon mass;
$m_{\tilde{\mu}_L}$ $(m_{\tilde{\mu}_R})$ is the mass of the left-handed  (right-handed) smuon;
$f(x,y)$ is a loop function given in \cite{Cho:2011rk}; $\Delta_\mu$ and $\delta_{\rm QED}$ are two-loop corrections given in~\cite{Marchetti:2008hw,Degrassi:1998es} which are of $\mathcal{O}(0.1)$.
In particular, the sign of the muon $g-2$ is correct if $\rm sign$$\mu>0$ which we assume here and hereafter.\footnote{It is difficult to explain the muon $g-2$ anomaly with $\mu<0$ and $1+\D_\mu <0$  due to the vacuum stability bound~\cite{Endo:2019bcj}.
The exception is $\D_\mu= -\O(10)$, which is hard to achieve in Higgs-anomaly mediation.  }

\subsection{ Muon $g-2$ with $M_{\rm inp}=10^{16}\GEV$}
To estimate the $g-2$, we perform the numerical simulation by using {\tt SuSpect\,2.4.3}~\cite{Djouadi:2002ze} with 
appropriate modifications: in particular, we take $y_b$ and $y_{\tau}$ to be large enough by hand at the early stage of iterations such that \eqref{eq:mA} is satisfied.
We take, throughout the paper, the top mass as $M_t=173.34\GEV$ and QCD 
coupling constant as $\alpha_s(m_Z)=0.1181.$ 
The numerical result of the muon $g-2$ and the mass spectrum is shown in the sample point $\bf I$ of 
Table.\ref{tab:1} by taking $M_{\rm inp}=10^{16}\GEV$ as in~\cite{Yin:2016shg, Yanagida:2016kag}. 
Here, $M_{\rm inp}=10^{16}\GEV$ corresponds to the case where the compactification scale ($1/L$) is around the GUT scale, and thus the MSSM with the SUSY breaking parameters of \eq{HM} appears at the scale. 
In fact, this sample point almost maximizes the $g-2$. 
This can be found from the facts that i) the current LHC bound on the wino-like neutralino mass 
$460\GEV$~\cite{Aaboud:2017mpt}, ii) the selectrons should be heavier than the wino 
to realize the wino-like neutralino as the lightest superpartner (LSP). 
The two conditions imply that the muon $g-2$ can not be increased by decreasing $m_{3/2}$ or $|c_h|$, which decreases wino mass or selectron masses from \eq{gaugino} or \eq{HM2}. 
In fact, the variation of $\tan\b$ causes the split between the selectron masses, and thus it also leads to the selectron LSP~\cite{Yanagida:2018eho}.  
Therefore, the explanation of the muon $g-2$ anomaly at the $1\s$ level with $M_{\rm inp}=10^{16}\GEV$ is driven into a corner and is excluded if we take the $g-2$ theoretical value from Ref.~\cite{Keshavarzi:2019abf}. This means that the brane-world scenario with compactification scale at $10^{16}\GEV$ can not explain the muon $g-2$ at $1\s$ level.

Before ending this section let us make a few discussions. 
We notice that the Higgs boson mass, estimated by {\tt FeynHiggs 2.14.0}~\cite{feynhiggs,feynhiggs2,feynhiggs3,feynhiggs4,feynhiggs5,feynhiggs6,feynhiggs7,feynhiggs8}, is slightly below the measured value~$125\GEV$. The Higgs boson mass, however, has a theoretical uncertainty larger than a few GeV from the measured top quark mass and higher order loop corrections (the dominant uncertainty comes from the resummation of the sbottom loops).\footnote{We thank Sven Heinemeyer for useful communication.} 
The second discussion is on the wino-like neutralino LSP. 
If the wino composes the dominant dark matter,\footnote{The thermal abundance is not enough to compose the dominant dark matter. We need non-thermal production of the dark matter e.g. from the gravitino decays. } it is disfavored from the null result of indirect detection experiments~\cite{Bhattacherjee:2014dya}, but is not excluded. 
Although we have light gluino and squarks, the current result of the multi-jet search with zero-lepton in the LHC~\cite{ATLAS:2019vcq} based on the simplified model can not be directly
applied since our spectrum is too different.  
In particular, most squarks decay to the bino and then the subsequent decay produces leptons. 
A detail collider simulation is important but beyond the scope of the present paper.

\section{Muon $g-2$ with $M_{\rm inp} \sim 10^{12}\GEV$}
\label{sec:3}

In this section, we show that the muon $g-2$ can be explained at the $1\s$ level with $M_{\rm inp} \sim 10^{12}\GEV$, which implies the dynamical generation of $m_{H_{u,d}}$ at an intermediate scale.
In fact, $M_{\rm inp}$ at the intermediate scale is theoretically possible. 
For not too small $M_{\rm inp}\lesssim 10^{16}\GEV$, we can still have the brane-world scenario with compactification scale, $1/L$, being $M_{\rm inp}$.
For $M_{\rm inp}\ll 10^{16}\GEV$, however, it is questioned whether the brane-world scenario with a small $1/L$ has a UV completion. 
On the other hand, there is a possibility with small $M_{\rm inp}$ but large compactification scale, $1/L \sim 10^{16}\GEV \gg M_{\rm inp}:$ the SUSY breaking masses for the Higgs doublets are dynamically generated at around $M_{\rm inp}.$ In this section, we take this possibility. A possible UV model is given in Sec.\ref{Sec:4}.

A smaller $M_{\rm inp}$ would effectively lead to larger $|\m|$-term and thus larger muon $g-2$ (see \eq{gm2}). 
To see this, let us fix the gravitino mass. This means the anomaly mediation contribution is almost fixed at the scale $\m_{\rm RG}\sim m_{3/2},$ and thus the gaugino masses are also fixed. (Remember that 
we can use the anomaly mediation formulae due to UV insensitivity.) If we decrease $M_{\rm inp}$ for fixed $c_h$, the smuon and selectron masses decrease due to the smaller logarithmic factor (see e.g. \eq{HM2}). 
However, one can increase the slepton masses by increasing $|c_h|$, which enlarges positive contributions from e.g.~\eq{HM2}. Thus, by choosing larger $c_h$, one can (almost) fix the smuon and gaugino masses while decreasing $M_{\rm inp}.$ Importantly, $|\mu|$-term is increased due to \eq{mu}, which enhances the muon $g-2$ in \eq{gm2}. 
 
\begin{figure}[!t]
\begin{center}  
   \includegraphics[width=80mm]{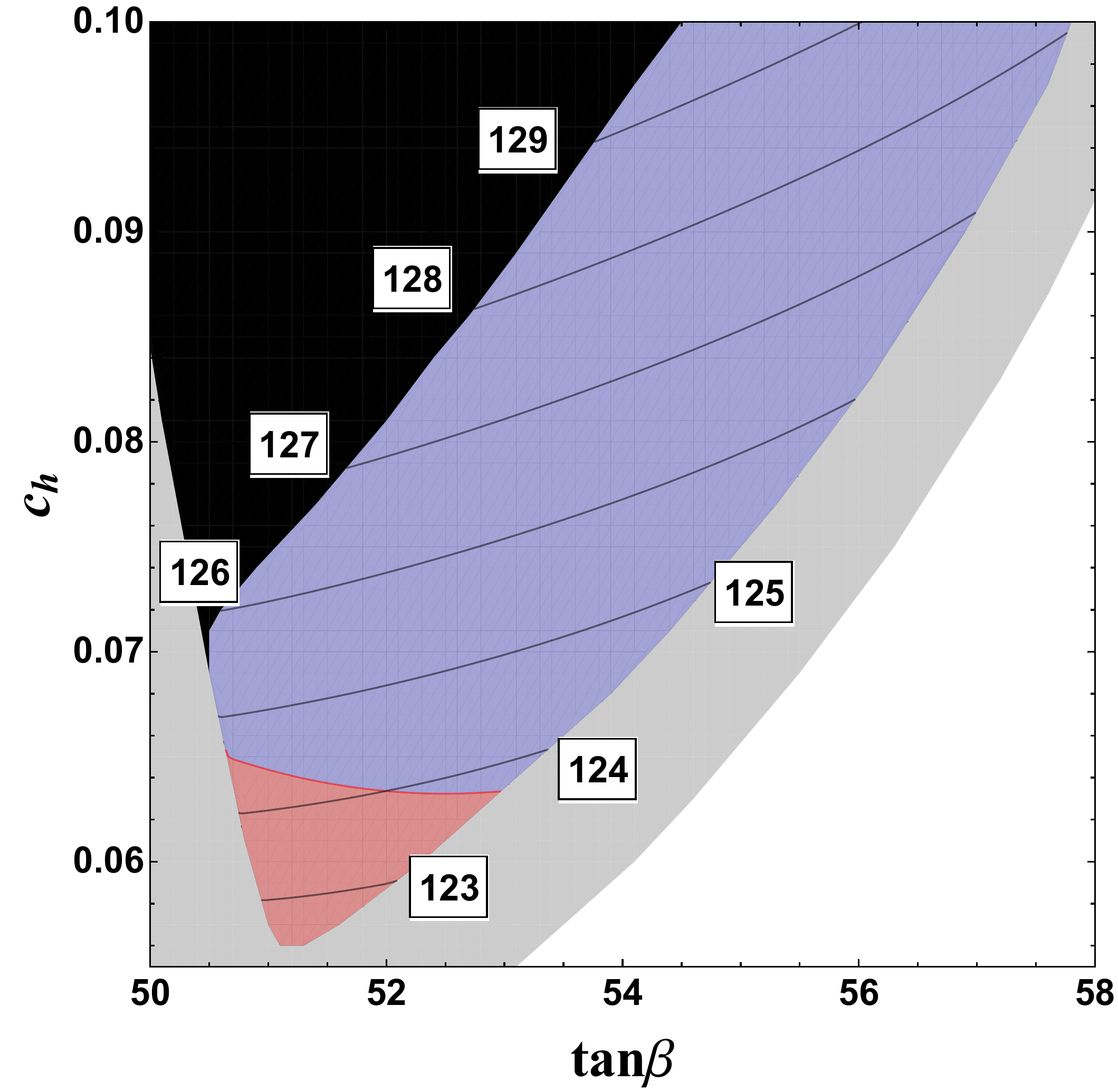}
   \includegraphics[width=80mm]{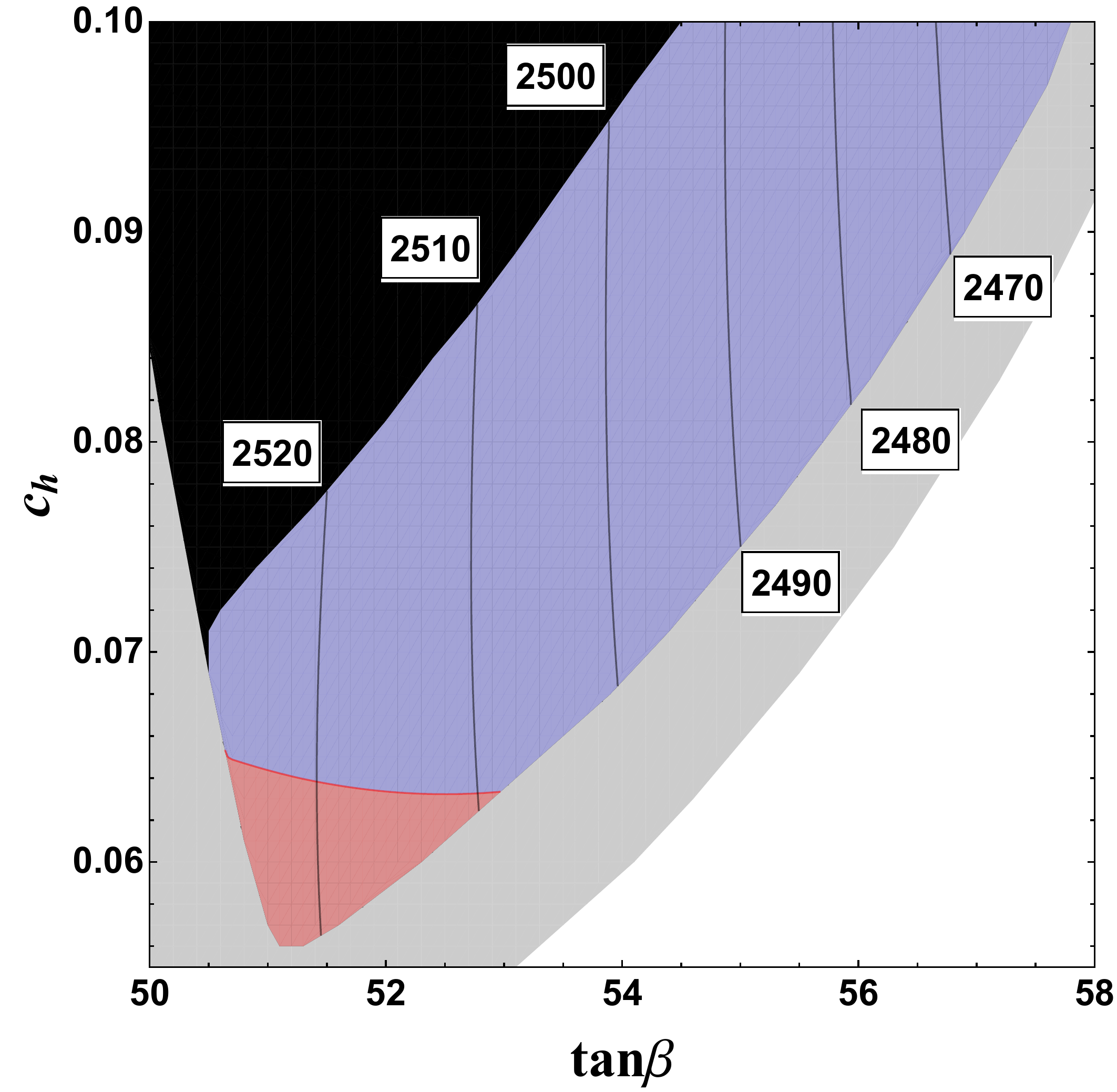}
\end{center}
\caption{The $1\sigma$ [red] and $2 \s$ [blue] regions of the muon $g-2$. The contours of the Higgs boson mass [GeV] and the left-handed squark mass [GeV] in first two generations are also shown in the left and right panels respectively. Black region may be excluded due to the vacuum decay. On the gray region the LSP is not the wino-like neutralino. 
We take $m_{3/2}=146\,\TEV, M_{\rm inp}=3\times 10^{12}\GEV$.
}
\label{fig:1}
\end{figure}

In Fig.\,\ref{fig:1}, we show the numerical results of the 1$\s$ (red band) and $2\s$ (blue band) regions of the muon $g-2$, the contours of Higgs boson mass (left-panel) and the averaged left-handed squark mass (right-panel) in first two generations for $M_{\rm inp}=3\times 10^{12}\GEV$ and $m_{3/2}=146\TEV$ are shown.  
The wino (gluino) mass is around $463\GEV$ ($3060\GEV$). In the gray region, the wino is not the LSP 
and the scenario is cosmologically inconsistent unless (small) $R$-parity violation is assumed. 
The black region may be excluded by the vacuum decay to a color/charge breaking one, setting a constraint~\cite{Kusenko:1996jn}, 
$
7.5(m_{\tl{Q}_3}^2+m_{\tl{u}_3}^2)>3\mu^2 +A_t^2.
$
This restricts $M_{\rm inp}\gtrsim 10^{12}\GEV,$ otherwise $|\mu|$ becomes too large to satisfy it. 
We find that the parameter region of the $1\sigma$ explanation of the $g-2$ is enlarged. 
Sample points {\bf II}\AND {\bf III} are shown in Table. \ref{tab:1}.

\begin{table*}[!t]
\caption{Mass spectra for some model points. On all the points the wino is the LSP.
 We denote the second and first generation squarks by $\tilde{X}_{(2,1)}$. For the mass of $\tilde{Q}_{(2,1)}$, 
 the average of the up and down type squark masses is taken. Here, $c_h$ and $\tan\beta$ are set at $M_{\rm inp}$ 
 and $\sqrt{ m_{\tl{t}_1} m_{\tl{t}_2}}$, respectively. The Point {\bf I} is out of the $1\sigma$ range of the muon $g-2$ if we adopt the result of Ref.~\cite{Keshavarzi:2019abf}.}
\label{tab:1}
\begin{center}
\begin{tabular}{|c||c|c|c|c|}
\hline
Parameters & Point {\bf I} & Point {\bf II }   & Point {\bf III} \\ 
\hline
$m_{3/2} $ (TeV) & 145.2  & 149.0& 146  \\
$c_h$  & $0.0401$  & $0.06$  & $0.06$\\
$\tan\beta$  & 50.67  & 51.00  & 51.00 \\
$\log_{10}[{M_{\rm inp}/\GEV}]$& 16.0  & 12.5 & $12.2$\\
\hline
Particles & Mass (GeV) & Mass (GeV)& Mass (GeV) \\
\hline
$\tilde{g}$ & 3020& 3110 & 3050 \\
$\tilde{\chi}_{1,2}^0$ & 460, 1330 & 472, 1370 & 463, 1340 \\
$\tilde{t}_{1,2}$ (TeV) & 12.3, 12.6 & 14.1, 14.6 & 13.6, 14.0 \\
$\tilde{b}_{1,2}$ (TeV) & 12.9, 13.5 & 15.0, 15.7 & 14.6, 14.3 \\
$\tilde{Q}_{(2,1)}$  &2500, 2480& 2580, 2560 & 2540, 2520\\
$\tilde{u}_{(2,1)}$ &2290, 2290& 2350, 2350 & 2320, 2320 \\
$\tilde{d}_{(2,1)}$ &2400, 2400& 2460, 2460 & 2430, 2420 \\
$\tilde{e}_{L, R}$ &469, 465& 550, 483& 549, 614\\
$\tilde{\mu}_{L,R}$ &526, 574& 598,  584& 579, 667\\
$\tilde{\tau}_{1,2}$ (TeV) & 8.22, 11.7& 8.91, 12.7& 8.60, 12.2\\
$H^\pm$\,(TeV) & 10.1 & 11.5 &11.3 \\
$\tl{\chi}^{\pm}_2$ (TeV) & 24.59 & 31.8  & 31.2 \\
$h_{\rm SM\mathchar`-like}$ & 123 &  124  & 123 \\
\hline
$10^{9}\d \a_\mu $& 2.03 &  2.06  & 2.25 \\
\hline
\end{tabular}
\end{center}
\end{table*}

\section{A model for $m_{H_u}^2 \simeq m_{H_d}^2 < 0$}
\label{Sec:4}
We consider a simple example model to generate $m_{H_u}^2 \simeq m_{H_d}^2 < 0$ at the intermediate scale.
The superpotential is given by~\footnote{This superpotential is the same as that in Ref.~\cite{Lu:2013cta}. However, the purpose here is completely different.}
\begin{eqnarray}
W = \lambda_S S H_u H_d + M_S S \bar S + c_2 m_{3/2} H_u H_d,
\end{eqnarray}
where $S$ and $\bar S$ are gauge singlet superfields. 
Here, $M_S$ is a free parameter, which can be naturally smaller than $10^{16}$\,GeV since a symmetry recovers in the limit $M_S \to 0$.
With the Yukawa interaction, $\lambda_S S H_u H_d$, and a non-tachyonic soft SUSY breaking mass for $S$, $m_S^2 \sim m_{3/2}^2$, the soft SUSY breaking masses for the Higgs doublets are generated at $M_S$ as
\begin{eqnarray}
m_{H_u}^2 \simeq m_{H_d}^2 \simeq -\frac{|\lambda_S|^2}{8\pi^2} m_S^2 \ln \frac{M_{*}}{M_S}, \label{eq:super}
\end{eqnarray}
with $M_{*}$ being the the energy scale where $m_S^2$ is generated. For instance, $M_*$ can be identified as the GUT scale. The SUSY mass parameter, $M_S$, can be roughly regarded as $M_{\rm inp}$: $\ln(M_S) \sim \ln(M_{\rm inp})$. Note that $m_{H_{u,d}}^2$ are negative with $m_S^2>0$. The soft SUSY breaking mass for $S$ is generated with the 
the K\"{a}hler potential: 
\begin{eqnarray}
K' = - 3 M_{P}^2 \ln \left[1 - \frac{f(Z, Z^\dag) + \phi_i^\dag \phi_i + |S|^2 + |\bar S|^2 + c_1 H_u H_d + h.c. + \Delta K'}{3 M_{P}^2} \right], \label{eq:kahler}
\end{eqnarray}
where 
\begin{eqnarray}
\Delta K' = c_S \frac{|Z|^2}{M_P^2} |S|^2,
\end{eqnarray}
with $c_S$ of $\mathcal{O}(1)$. 
The setup can be justified in a brane world scenario with $1/L \sim 10^{16}\GEV$, where $S$ lives in the bulk and the MSSM fields including $H_u$ and $H_d$ live in the visible brane.\footnote{The singlet field $\bar S$ can live either in the bulk or the visible brane. In the former case, $\bar S$ is also expected to have a soft SUSY breaking mass, which does not affect the generation of $m_{H_{u,d}}$.}
From \eqref{eq:super} and \eqref{eq:kahler}, the $\mu$-term and Higgs $B$-term are 
\begin{eqnarray}
\mu &=& (c_1 + c_2) m_{3/2}, \nonumber \\
B\mu &=& (c_1-c_2) m_{3/2}^2.
\end{eqnarray}
From the conditions of EWSB with $\tan\beta=\mathcal{O}(10)$, $c_1 + c_2 = \mathcal{O}(0.1)$ and $c_1 - c_2 = \mathcal{O}(10^{-3})$ should be satisfied: to explain the correct weak scale, we need a fine-tuning of $\mathcal{O}(10^{-2})$. This fine-tuning for the small $B$-term is a common problem in scenarios with $m_{3/2}=\mathcal{O}(100)$\,TeV and $\tan\beta={O}(10)$.

The consistent charge assignment is given in Table~\ref{tab:2}.\footnote{
This charge assignment is consistent with the seesaw mechanism~\cite{Yanagida:1979as, GellMann:1980vs, Glashow:1979nm} (see also Ref.~\cite{Minkowski:1977sc}) with the $U(1)_R$ charges for $Q,u,d,L,e$ and $\bar N$ being one. Here, $\bar N$ is a chiral multiplet of right-handed neutrino.
} With this charge assignment, we may also have
\begin{eqnarray}
W = \xi_S S + c_3 m_{3/2} M_S \bar S.
\end{eqnarray}
The above terms do not generate too large $\left<S\right>$ and $\left<F_S\right>$ (and hence $\mu$ and $B\mu$) when the conditions, $\xi_S < \mathcal{O}(0.1) M_S^2$ and $c_3 < \mathcal{O}(0.1)$, are satisfied:\footnote{These conditions are required to explain the correct EWSB.
Alternatively, instead of $U(1)_{\bar{S}}$, one can consider the Peccei-Quinn symmetry, under which $S$, $\bar S$ and $H_u$ are charged (see Appendix A). With the symmetry, $c_{1,2}$ are suppressed but still one can have successful EWSB with the VEVs of $S$ and $F_S$. 
} $\left<S\right> \simeq (\xi_S/M_S^2 -c_3) m_{3/2} $ and $\left<F_S\right>=-(\xi_S/M_S^2+c_3) m_{3/2}^2$.\footnote{Here, $\mathcal{L} = \int d^2 \theta (\Phi M_S S \bar S + \Phi^2 \xi_S S + c_3 \Phi^2 m_{3/2} M_S \bar S) + h.c.$ is used, where $\Phi=1+m_{3/2} \theta^2$ is the conformal compensator field.}

\begin{table*}[!t]
\caption{\small Charge assignment}
\label{tab:2}
\begin{center}
\begin{tabular}{c||c|c|c|c|c|c}
Symmetry & $S$ & $\bar S$ & $H_u$ & $H_d$ & $m_{3/2}$ & $M_S$ \\
\hline
$U(1)_R$ & 2 & 0 & 0 & 0 & 2 & 0   \\ 
\hline
$U(1)_{\bar S}$  & 0 &  1 & 0 & 0 & 0 & -1 
\end{tabular}
\end{center}
\end{table*}

\section{Conclusions and discussions}

In this paper, we have explored the $1\s$ explanation of the muon $g-2$ in Higgs-anomaly mediation. In Higgs-anomaly mediation, squarks and sleptons are massless at the tree level at a high energy scale such as the GUT scale while the Higgs doublets get large and negative SUSY breaking masses squared at a certain energy scale, $M_{\rm inp}$.
The masslessness of squarks and sleptons arising from the sequestered K\"{a}hler potential is a simple and important assumption to avoid the SUSY flavor problem while respecting the GUT symmetry. (Remember that the squarks and sleptons live in a same GUT multiplet.) The SUSY breaking field is assumed to carry some conserved charge to avoid the Polonyi problem.  
In this case, gaugino masses are also vanishing at the tree level and they are solely determined by anomaly mediation. The masses for squarks and sleptons are radiatively generated from anomaly mediation and Higgs-loop effects. The Higgs loop effects increase the masses for third generation sfermions significantly, which explains the observed Higgs boson mass, and solves the tachyonic slepton problem in the originally proposed anomaly mediation scenario. The smuon and bino masses are small enough to enhance the muon $g-2$. Moreover, the gravitino problem is greatly relaxed due to the heavy gravitino $m_{3/2}\gtrsim 100\TEV$.

We have found that, from the latest LHC data, the $1\s$ level explanation of the muon $g-2$ suggests $M_{\rm inp} \sim 10^{12}\GEV$. This implies that $m_{H_u}^2 \simeq m_{H_d}^2 <0$ are generated dynamically at the intermediate scale. 
In Sec.~\ref{Sec:4}, we provide a simple model to realize $m_{H_u}^2 \simeq m_{H_d}^2 <0$ at the intermediate scale using $\lambda_S S H_u H_d$ while keeping the sequestering scale to be around the GUT scale. We note that, since $m_{H_{u,d}}$ is generated through the renormalization group evolution from the GUT scale to the SUSY mass for $S$, the mass spectrum of MSSM particles may be slightly different from that in the scenario where $m_{H_u}^2 \simeq m_{H_d}^2$ is set at $M_{\rm inp}$ by hand. The detailed study will be discussed elsewhere. 

The wino is the LSP in our scenario with the mass in the sub-TeV range. This wino can be tested at the LHC by looking for disappearing charged tracks. Alternatively, the wino dark matter may be tested by indirect detection experiments in a few years. There are also various light sfermions which are fully covered at LHC and future collider experiments. 

\section*{Acknowledgments}
We greatly thank Sven Heinemeyer for communication on {\tt FeynHiggs} calculation of the Higgs boson mass. 
T. T. Y. is supported in part by the China Grant for Talent Scientific Start-Up Project and the JSPS Grant-in-Aid for Scientific Research No. 16H02176, and No. 17H02878, No. 19H05810 and by World Premier International Research Center Initiative (WPI Initiative), MEXT, Japan.
W.Y. is supported by NRF Strategic Research Program NRF-2017R1E1A1A01072736. N.Y. is supported by JSPS KAKENHI Grant Numbers JP15H05889, JP15K21733, and JP17H02875.

\appendix

\section{A model with PQ symmetry}
\label{sec:PQM}
In this appendix, we present another model to generate $m_{H_u}^2 \simeq m_{H_d}^2 <0$ at the intermediate scale, with successful EWSB. The model has the Peccei-Quinn (PQ) symmetry and charge assignments are given in Table \ref{tab:ap}. The superpotential is 
\begin{eqnarray}
W = \lambda_S S H_u H_d + k_1 M_S^2 S + M_S S \bar{S} + m_{3/2} \bar{M} \bar{S} + \kappa S \bar{S}^2,
\end{eqnarray}
where $M_S$ and $\bar{M}$ are spurious fields breaking $U(1)_{\rm PQ}$, and $M_S \sim \bar{M}$.
As in the model presented in Sec.~\ref{Sec:4}, $m_{H_u}^2 \simeq m_{H_d}^2 < 0$ are generated by $S H_u H_d$ term. With the PQ symmetry, we can not write bare $\mu$-term and $B_\mu$-term. However, they are generated by the VEVs of $S$ and $F_S$. By minimizing the potential, we obtain
\begin{eqnarray}
\left<S\right> &\simeq& (k_1-k_2)m_{3/2}+\kappa(3 k_1^2 - 2 k_1 k_2) m_{3/2}, \nonumber \\
\left<F_S\right> &\simeq& -(k_1+k_2)m_{3/2}^2 + \kappa(-3k_1^2-2 k_1 k_2) m_{3/2}^2, 
\end{eqnarray}
where we have neglected higher order terms of $\kappa$ and $m_{3/2}$, and $k_2 = \bar{M}/M_S$. Note that for $k_1 \sim k_2 = \mathcal{O}(0.1)$ and $k_1 + k_2 =  \mathcal{O}(10^{-3})$, one can obtain the desired sizes of $\mu$-term and $B_\mu$-term.

\begin{table*}[!t]
\caption{\small Charge assignment}
\label{tab:ap}
\begin{center}
\begin{tabular}{c||c|c|c|c|c|c|c}
Symmetry & $S$ & $\bar S$ & $H_u$ & $H_d$ & $m_{3/2}$ & $M_S$ & $\bar{M}$\\
\hline
$U(1)_R$ & 2 & 0 & 0 & 0 & 2 & 0 & 0 \\ 
\hline
$U(1)_{\rm PQ}$ & $2$ & $-1$ & $-2$ & $0$ & 0 & $-1$ & $1$ \\
\end{tabular}
\end{center}
\end{table*}

\providecommand{\href}[2]{#2}\begingroup\raggedright\endgroup

\end{document}